\documentclass[10pt,letterpaper,twocolumn,superbib,pra]{revtex4}
\usepackage{amssymb}
\usepackage{amsmath}
\usepackage{graphicx}

\input{tcilatex}

\begin{document}

\title{Cavity solitons in bidirectional lasers}
\author{Isabel P\'{e}rez--Arjona, V\'{\i}ctor J. S\'{a}nchez--Morcillo, }
\affiliation{Departament de F\'{\i}sica Aplicada, Escola Polit\`{e}cnica Superior de
Gandia, Universitat Polit\`{e}cnica de Val\`{e}ncia, Ctra. Nazaret--Oliva
S/N, 46730--Grau de Gandia, Spain}
\author{and Eugenio Rold\'{a}n}
\affiliation{Departament d'\`{O}ptica, Universitat de Val\`{e}ncia, Dr. Moliner 50,
46100--Burjassot, Spain}

\begin{abstract}
We show theoretically that a broad area bidirectional laser with slightly
different cavity losses for the two counterpropagating fields sustains
cavity solitons (CSs). These structures are complementary, i.e., there is a
bright (dark) CS in the field with more (less) losses. Interestingly, the
CSs can be written/erased by injecting suitable pulses in any of the two
counterpropagating fields.
\end{abstract}

\maketitle

Cavity solitons are localized structures that form in broad area nonlinear
optical cavities in the plane orthogonal to the field propagation direction.
They are interesting not only from a fundamental point of view but also
because their potentialities in optical information storage and processing 
\cite{Staliunas02}.

Interestingly, the most important nonlinear optical cavity, namely the
laser, does not exhibit CSs easily, i.e., something must be done for usual
lasers display localized structures different from light vortices. One
possibility is introducing a suitable saturable absorber in the laser cavity 
\cite{Rosanov96}. Another possibility consists in \textit{rocking} the laser 
\cite{Valcarcel03}, i.e., in injecting an amplitude modulated signal. These
two ways have been experimentally demonstrated in lasers \cite{Taranenko97}
or laser--like systems \cite{Esteban06}. Apart from these, it has been
numerically shown that cascade lasers \cite{Vilaseca01}, and lasers with a
dense active medium \cite{Ahufinger03}, can sustain CSs (in the second case
some intracavity spatial filtering must be done), but up to now there is no
experimental evidence for these predictions.

In this letter we numerically show that CSs can form in bidirectional lasers
provided that the two counterpropagating fields experience slightly
different losses (up to a few per cent). Interestingly, CSs in the two
fields are complementary (a bright CS in one field complements with a dark
CS\ in the other field) and can be written/erased by acting on only one of
the two fields.

Consider a broad--area ring--cavity filled with a homogeneously broadened
two--level active medium which is incoherently pumped. The Maxwell--Bloch
equations of this system, within the uniform field limit and rotating
wave--approximations, are \cite{Zeghlache}

\begin{subequations}
\label{model}
\begin{eqnarray}
\partial _{t}E_{n} &=&-\kappa _{n}E_{n}-i\,A\,\left\langle
P_{n}\right\rangle +i\,a\bar{\nabla}_{\bot }^{2}E_{n}, \\
\partial _{t}P_{n} &=&-\gamma _{\bot }(1-i\Delta )P_{n} \\
&&+i\gamma _{\bot }D\left\{ E_{n}+E_{m}\exp \left[ \left( -1\right) ^{n}2ikz%
\right] \right\}  \notag \\
\partial _{t}D &=&-\gamma _{\Vert }\left[ D-1-\func{Im}\left(
E_{1}P_{1}^{\ast }+E_{2}P_{2}^{\ast }\right) \right] ,
\end{eqnarray}%
with $n,m=1,2$ and $n\neq m$. In the above equations $E_{n}\left(
x,y,t\right) $, denote the field amplitudes of the two ($n=1,2$)
counterpropagating fields traveling along the $z$ direction, $P_{n}\left(
x,y,z,t\right) $ are their respective medium polarization sources, $D\left(
x,y,z,t\right) $ denotes the population inversion, and 
\end{subequations}
\begin{equation}
\,\left\langle P_{n}\left( x,y,t\right) \right\rangle =\frac{1}{L_{z}}%
\dint\limits_{0}^{L_{z}}dz\,P_{n}\left( x,y,z,t\right) ,
\end{equation}%
are the axially--averaged medium polarizations with $L_{z}$ the laser cavity
length. $\bar{\nabla}_{\bot }^{2}=$ $\left( \partial _{x}^{2}+\partial
_{y}^{2}\right) $ is the transverse Laplacian and regarding the parameters, $%
\Delta =\left( \omega _{C}-\omega _{at}\right) /\gamma _{\bot }$ is the
normalized cavity detuning ($\omega _{C}$ and $\omega _{at}$ are,
respectively, the cavity and atomic transition frequencies), $A$ is the pump
parameter, $a$ is the diffraction coefficient, and the field, medium
polarization and inversion decay rates are denoted by $\kappa _{n}$, $\gamma
_{\bot }$, and $\gamma _{||}$, respectively.

In order to keep the mathematical modelling as simple as possible, we shall
restrict our study here to class--A lasers (i.e., those verifying $\gamma
_{\bot }\gg \gamma _{||}\gg \kappa _{1},\kappa _{2}$) and shall consider
perfect resonance, i.e., $\Delta =0$. For class--A lasers all material
variables can be adiabatically eliminated. After taking $\partial
_{t}P_{n}=\partial _{t}D=0$, one easily gets 
\begin{equation}
P_{n}=\frac{i\left\{ E_{n}+E_{m}\exp \left[ \left( -1\right) ^{n}2ikz\right]
\right\} }{1+\left\vert E_{1}\right\vert ^{2}+\left\vert E_{2}\right\vert
^{2}+E_{1}E_{2}^{\ast }e^{2ikz}+c.c.}.  \label{aux}
\end{equation}%
Now we assume that the pump parameter value is close to its value at lasing
threshold, so that field amplitudes are small quantities. Then, after Taylor
expanding Eq. (\ref{aux}), it follows that $\left\langle P_{n}\right\rangle
=iE_{n}\left( 1-\left\vert E_{n}\right\vert ^{2}-2\left\vert
E_{m}\right\vert ^{2}\right) $, and then Eqs. (\ref{model}) reduce to 
\begin{subequations}
\label{GL}
\begin{eqnarray}
\partial _{\tau }F_{1} &=&\left( A-1-\,\,\left\vert F_{1}\right\vert
^{2}-2\left\vert F_{2}\right\vert ^{2}\right) F_{1}+i\,\nabla _{\bot
}^{2}F_{1}, \\
\partial _{\tau }F_{2} &=&\left( A-\sigma -\,\,\left\vert F_{2}\right\vert
^{2}-2\left\vert F_{1}\right\vert ^{2}\right) F_{2}+i\,\nabla _{\bot
}^{2}F_{2},
\end{eqnarray}%
where we have introduced $F_{n}=\sqrt{A}E_{n}$, $\nabla _{\bot }^{2}=a\bar{%
\nabla}_{\bot }^{2}$, $\tau =\kappa _{1}t$, and $\sigma =\kappa _{2}/\kappa
_{1}$ being $\xi =x/\sqrt{a}$ the dimensionless transverse coordinate. This
is the model for a resonant class--A bidirectional laser we shall study
here, which consists of a pair of coupled Ginzburg--Landau equations. It is
worth mentioning that the same equations are obtained by performing a
rigorous derivation based on the multiscale expansion technique, which we
shall publish elsewhere. Notice that Eqs. (\ref{GL}) have only two
independent parameters: pump $A$ and quotient of cavity losses $\sigma $,
which we take $\sigma \geq 1$ without loss of generality.

Eqs. (\ref{GL}) have four possible stationary homogeneous solutions: The
laser-off solution, $F_{1}=F_{2}=0$, the \textit{strong} unidirectional (or
singlemode) solution $\left\{ F_{1}=\sqrt{A-1},F_{2}=0\right\} $, the 
\textit{weak} unidirectional solution $\left\{ F_{1}=0,F_{2}=\sqrt{A-\sigma }%
\right\} $, and the bidirectional (or bimode) solution $\left\{ F_{1}=\left(
A-2\sigma +1\right) /3,F_{2}=\left( A-2+\sigma \right) /3\right\} $. The
linear stability analyses of these solutions shows that: (i), The laser--off
solution losses its stability at $A=1$ giving rise to the strong
unidirectional solution, which is always linearly stable; (ii), the weak
unidirectional solution is linearly stable for $A\geq 2\sigma -1$; and
(iii), the bidirectional solution is always unstable. Thus bidirectional
homogeneous cw emission is impossible, as it is well known from the
plane--wave theory \cite{Zeghlache}. Our basic hypothesis is that stable
bidirectional cw emission could be possible in a nonhomogeneous solution and
this is the possibility we examine now through the numerical integration of
Eqs. (\ref{GL}) with the purpose of finding CS solutions.

We shall concentrate here in the one--transverse dimension limit, i.e., we
assume that diffraction along one transverse direction is avoided by some
means, while it acts freely on the other (orthogonal) coordinate (think,
e.g., of a waveguide configuration in one transverse direction). We have
used a split--step algorithm for solving Eqs. (\ref{GL}) with a grid of 512
spatial points and a total integration length $L/\sqrt{a}=1$ with $a=10^{-5}$%
. In order to approach realistic conditions, here we consider that pump is
limited in the transverse direction and further assume a top--hat like
profile for $A$ (for computational convenience we take $A\left( \xi \right)
=A\exp \left[ -\left( \xi /\xi _{p}\right) ^{2n}\right] $ with $\xi
_{p}=0.45 $ and $n=10$). Nevertheless, we have also carried out integrations
with an unlimited pump function in order to check the robustness of our
results.

We have first investigated the dynamics of the system when a real positive
pulse is injected in the weak mode (its maximum value equals the weak--mode
homogeneous solution), and the strong--mode homogeneous solution, with a dip
arriving to zero in its center, is injected in the strong mode. Remarkably,
the results are independent of the exact shape and width of the pulse (which
are the same as that of the dip).

For $\sigma =1$ this initial condition relaxes to a final state in which the
weak mode is null in all the space irrespective of the pump value \cite%
{coment1}. A quite different result is reached as $\sigma $ is made
different from unity and $A\geq 1.05$: a narrow domain of emission in the
weak mode, and a corresponding dip in the emission of the strong mode,\ are
formed where the initial condition was injected, giving rise to a localized
structure.

In Fig. 1 we show, in the $\left\langle \sigma ,A\right\rangle $ plane, the
domain of existence of these localized structures. In the inset an example
of the intensity profiles of the two fields is shown. The bright CS of the
weak mode is nearly Gaussian, while the dark CS --or, more precisely, the
grey CS, as the minimum intensity is clearly different from zero-- of the
strong mode corresponds to a nearly supergaussian dip on the strong--mode
homogeneous solution. The CSs represented in the inset of Fig. 1 correspond
to the domain in which they are stationary but the existence of a Hopf
bifurcation as $\sigma $ is decreased must be noticed. The Hopf bifurcation
is supercritical, i.e. when reached (decreasing $\sigma $) small
oscillations appear that affect both the height and width of the CS,
oscillations that grow in amplitude as $\sigma $ is further decreased. In
Fig. 2 an example of the dynamic behavior exhibited by the CS is shown. Fig.
2(a) shows the spatiotemporal oscillation of the two fields, the maximum of
weak field, $F_{2}$, exhibits periodic oscillations, see Fig. 2(b).

The continuous line in Fig. 1 corresponds to the boundary where the CSs
cease to exist. When crossing this line, the bright CS of the weak mode
simply decreases in amplitude, quickly, till it disappears, giving rise to
homogeneous emission in the strong mode. Something similar occurs at $\sigma
=1$, where the oscillating CSs collapses.

After establishing the existence of these CSs, an important issue is to
determine whether they can be manipulated, i.e., how can they be
written/erased. Notice that in the bidirectional laser one can inject
signals in either the strong or the weak mode or in both. The results we
have just commented are obtained from an initial condition which, to some
extent, can be understood as the simultaneous injection of a positive pulse
in the weak mode and a negative pulse in the strong one. But from an
experimental perspective this could be difficult to implement. Moreover, one
must take into account the unavoidable presence of noise in real devices.

We have thus carried out a series of numerical integrations in which we have
tested the writing and erasing of CSs by injecting appropriate (i.e., with
the right phase) pulses in only one of the two counterpropagating fields in
the presence of additive noise (random noise with amplitude $10^{-5}$) in
Eqs. (\ref{GL}). The results are summarized in Fig. 3. The initial condition
consists of emission only in the strong mode. We start with homogeneous
steady state emission in the strong mode. At $\tau =0$, a positive Gaussian
pulse is added only in the weak mode for a short time, the result being the
formation of the CSs. The CSs are then removed by injecting a positive
Gaussian in the strong mode, at $\tau =2$\textperiodcentered $10^{4}$. Then,
at $\tau =2.5$\textperiodcentered $10^{4}$ a negative Gaussian is injected
in the strong mode, and the CSs are formed again, which are later removed by
injecting again at $\tau =3.5$\textperiodcentered $10^{4}$ a Gaussian pulse
in the weak field but with the opposite phase (the injection times are
marked in the figure with a horizontal white line). These numerical results
prove the robustness and addresability of these localized structures.

We have also assessed the possibility that the system supports clusters
formed by an array of CSs. The numerical results ensured that several CSs
can be switched on and that this solution is stable and robust in the
presence of noise. Figure 4 shows the spatiotemporal evolution where a
cluster formed by four CSs is developed (a)\ from the strong mode emission
initial condition when four Gaussian pulses are injected in the weak field
and the final intensity profile (b).

In conclusion we have shown that cavity solitons form in class--A
bidirectional lasers when the two counterpropagating fields experience
slightly different cavity losses. We find that it is particularly remarkable
the fact that (i) the CSs in the two fields are complementary, i.e., when
one field exhibits a bright CS, the counterpropagating field shows a dark
CS\ in the same position, and (ii)\ the CSs can be written/erased by acting
on only one of the two fields.

This work has been supported by the Spanish Ministerio de Educaci\'{o}n y
Ciencia and the European Union FEDER through Projects FIS2005-07931-C03-01
and -02 and Programa Juan de la Cierva.

\newpage \bigskip {\large FIGURE CAPTIONS}\bigskip

\textbf{Fig.1.} Bifurcation diagram of CSs in the $<\sigma ,A>$ plane.
Continuous and dashed lines represent the boundaries of existence of
stationary and dynamic CSs, respectively. The inset shows a stationary CS
obtained for $\sigma =1.01$ and $A=1.4$. Continuous (dashed) line
corresponds to the field amplitudes in the weak (strong) modes.

\textbf{Fig. 2.} Pulsing CS obtained for $\sigma =1.005$ and $A=1.4$. The
spatiotemporal evolution of the pulsing CS is shown in (a), where $\tau $
runs from $2.9~10^{5}$ to $3~10^{5}$ and $\xi $ from $-0.5$ to $0.5$, the
maximum of the CS in the weak field oscillates periodically in time (b).

\textbf{Fig. 3.} Switching on and off of CSs by different techniques.
Parameters are $\sigma =1.015$ and $A=1.4$, $\xi $ running from $-0.5$ to $%
0.5$. The white lines indicate the times at which pulses are injected. For
details, see text.

\textbf{Fig. 4.} Cluster formed by four CSs for $\sigma =1.005$ and $A=1.4$.
Time $\tau $ runs from $0$ to $2.5~10^{4}$. Spatiotemporal evolution (a)
from the injection and final intensity profile (b).
\end{subequations}


\begin{thebibliography}{9}
\bibitem{coment1} When the pump transverse extension is unlimited, the
dynamics is different:\ one observes that two complementary domains of
emission are formed in the two fields (where one mode is on the other mode
is off and viceversa) that oscillate expanding and contracting periodically.
This behavior disappears when the pump transverse extension is finite.
\end{thebibliography}
\end{document}